\documentclass[twocolumn,aps,prd,amsmath,amssymb,nofootinbib]{revtex4-1}\newcommand{\preprintnumber}{\hfill \,\,\,\,\maketitle}
\pdfoutput=1

\usepackage{booktabs}
\usepackage[english]{babel}
\usepackage{amsmath,amssymb,amsbsy,amstext, amsthm, simplewick,slashed}
\usepackage{graphicx}
\usepackage{amsfonts}
\usepackage{tikz}
\usepackage{upgreek}
\usepackage{simplewick}
\usepackage{exscale,relsize}
\usepackage{bbding}
\usepackage{tabu}
\usepackage{multirow}
\usepackage{color}
\usepackage{graphicx}  
\usepackage{dcolumn}   
\usepackage{bm}        
\usepackage{amssymb}   
\usepackage{amsmath} 
\usepackage{enumerate}
\usepackage{graphicx}  
\usepackage{dcolumn}   
\usepackage{bm}        
\usepackage{amssymb}   
\usepackage{amsmath} 
\usepackage{enumerate}

\usepackage{tikz}
\usetikzlibrary{decorations.pathmorphing,decorations.markings}

\tikzset{
photon/.style={decorate, decoration={snake,amplitude=4pt, segment length=7pt}, draw=black},
particle/.style={draw=black, postaction={decorate}, decoration={markings,mark=at position .5 with {\arrow[draw=black]{>}}}},
antiparticle/.style={draw=black, postaction={decorate}, decoration={markings,mark=at position .5 with {\arrow[draw=black]{<}}}},
gluon/.style={decorate, draw=black, decoration={coil,amplitude=3pt, segment length=4pt}},
higgs/.style={draw=black,dashed,thick },
arrow/.style={draw=black, very thick, postaction={decorate}, decoration={markings,mark=at position 1 with {\arrow[draw=black]{>}}}}
}

\newcommand{\mpl}{m_{\rm Pl}}

\definecolor{darklightsabergreen}{rgb}{0.0, .49, 0.06}

\begin{document}

\title{Clearing the Brush: The Last Stand of Solo Small Field Inflation}

\author{Joseph Bramante}
\affiliation{Department of Physics, University of Notre Dame, \\225 Nieuwland Science Hall, Notre Dame, IN 46556, USA}

\author{Sean Downes}
\affiliation{Leung Center for Cosmology and Particle Astrophysics, National Taiwan University, \\ No. 1, Sec. 4, Roosevelt Road, Taipei 10617, Taiwan}

\author{Landon Lehman}
\affiliation{Department of Physics, University of Notre Dame, \\225 Nieuwland Science Hall, Notre Dame, IN 46556, USA}

\author{Adam Martin}
\affiliation{Department of Physics, University of Notre Dame, \\225 Nieuwland Science Hall, Notre Dame, IN 46556, USA}

\begin{abstract}
By incorporating both the tensor-to-scalar ratio and the measured value of the spectral index, we set a bound on solo small field inflation of $\Delta \phi / \mpl \geq 1.00 \sqrt{r/0.1}$. Unlike previous bounds which require monotonic $\epsilon_V,$ $|\eta_V| <1$, and 60 e-folds of inflation, the bound remains valid for non-monotonic $\epsilon_V$, $|\eta_V| \gtrsim 1$, and for inflation which occurs only over the 8 e-folds which have been observed on the cosmic microwave background. The negative value of the spectral index over the observed 8 e-folds is what makes the bound strong; we illustrate this by surveying single field models and finding that for $r \gtrsim 0.1$ and 8 e-folds of inflation, there is no simple potential which reproduces observed CMB perturbations and remains sub-Planckian. Models that are sub-Planckian after 8 e-folds must be patched together with a second epoch of inflation that fills out the remaining $\sim 50$ e-folds. This second, post-CMB epoch is characterized by extremely small $\epsilon_V$ and therefore an increasing scalar power spectrum. Using the fact that large power can overabundantly produce primordial black holes, we bound the maximum energy level of the second phase of inflation.
\end{abstract}

\preprintnumber

\section{Introduction}
\label{sec:intro}

In this paper we find that if the BICEP2 result \cite{Ade:2014xna} indicating a large tensor-to-scalar ratio is confirmed, then sub-Planckian single field inflatons are severely restricted. While sub-Planckian inflatons are not technically extinct for $r \gtrsim 0.05$, the remaining examples will range over $\Delta \phi \gtrsim 0.7 ~ \mpl$ and are ``sub-Planckian" in name only. Moreover, we find that these barely sub-Planckian models must splice together two qualitatively different periods of inflation: one with the inflaton rolling quickly enough to generate tensor modes, and one with the inflaton rolling extremely slowly to remain sub-Planckian while generating the rest of inflation.

It is customary to motivate a period of cosmological inflation by pointing out problematic levels of isotropy and homogeneity in our universe. The well-established primordial epochs of matter and radiation dominated expansion, which are evident in big bang nucleosynthesis and the uniform temperature of the cosmic microwave background (CMB), funnel back in time to a singular conclusion: the universe is too flat and causally connected to have expanded throughout its history with a scale factor proportional to $t^n$ with $n > 0$. These flatness and horizon problems can be solved with a period of negative pressure that feeds an exponential expansion of the early universe, during which the universe's scale factor $a$ grows as $e^{Ht}$ where $H \equiv \dot{a}/a$. In order to be observed, such a period of exponential expansion must end. To this end we employ a quantized scalar field. The potential of a slowly rolling scalar field weakly breaks the time symmetry of exponentially expanding de Sitter space, making it quasi de Sitter, and thus permits the end of inflation. 

Quantized scalar fields vary as $\phi \rightarrow \phi + \delta \phi$, and so the variation of a de Sitter-sourcing action creates primordial scalar perturbations. Here the action being perturbed with respect to scalar degrees of freedom is the Einstein-Hilbert action for a scalar field, $S = \int \sqrt{-g} [\frac{1}{2}R+\frac{1}{2}g^{\mu \nu} \partial_\mu \phi \partial_\nu \phi -V(\phi)]$. The primordial scalar perturbations thus obtained have been observed in the different path lengths of photons in the cosmic microwave background and in the large scale distribution of matter in galactic surveys. Tensor perturbations of the Einstein-Hilbert action, on the other hand, do not depend on the form of the scalar potential, but rather on its magnitude. That is, they are set by the logarithmic rate of expansion of the universe $H = \delta_t ({\rm log}~a)$. A typical scalar potential is invariant under tensor perturbations, and the primordial tensor perturbation amplitude is solely a function of the energy density of de Sitter space, linked to the expansion rate of the de Sitter horizon. From the radiative perspective, the de Sitter horizon sources gravitons which become gravity waves, and the universe acts as a giant gravity wave interference plate, with photons Thomson-scattered off recombining globs of charge density as its signal.
\vskip1ex
If the scalar potential sources the vacuum energy of de Sitter space and $\frac{\mpl^2V_\phi^2}{2 V^2} \equiv \epsilon_V \ll 1$, where $V_\phi \equiv d V/d \phi$, this is called slow-roll inflation. Often it is required that $\frac{\mpl^2 V_{\phi \phi}}{V} \equiv \eta_V \ll 1$, but this is somewhat extraneous \cite{Dvorkin:2009ne}, as we will discuss. 

There are two ways to achieve inflation in a slow-roll potential ($\frac{\mpl^2V_\phi^2}{2 V^2} \ll 1$): the potential can be ``Planck flat" or ``Planck fat", more often identified as small or large field inflation. The field trajectory is Planck flat if the smallness of $V_\phi$ drives inflation. The smaller $V$ is with respect to the Planck scale, the smaller $V_\phi$ must be for successful Planck flat inflation. Planck fat inflation, on the other hand, is driven by a roughly Planckian field excursion during inflation: $\Delta \phi \sim \mpl$. We note that it is customary to define the value of $\phi$ as $\phi_*$ at the scale where CMB measurements ($r$, $n_s-1$, $A_s$) are centered, which is the pivot scale, and to define $\phi_e$ as the field value at the end of inflation. This leads to $\Delta \phi \simeq |\phi_* - \phi_e|$, where here we indicate a near equality, because a typical pivot scale $k_* \sim 0.002 ~{\rm Mpc^{-1}}$ is a little smaller than the Hubble scale $k_H=H_0\sim 10^{-4} ~{\rm Mpc^{-1}}$.

There are theoretical considerations which must be addressed for Planck fat and Planck flat inflation. It is not always possible to calculate perturbative corrections to Planck fat inflation. Consider the term $\frac{c |\phi|^5}{\Lambda}$. For an order one coupling constant $c$, and a Planck field range of $\Delta \phi \sim \mpl$, a ``sensible" cutoff scale must be super-Planckian. Given this, one might argue that Planck fat inflation is not describable with a normal effective field theory. To the contrary, one could argue that $\phi$ simply must be weakly coupled to any new dynamics and all standard model fields. For example, a simple realization of Planck fat inflation, e.g. $V = m^2 \phi^2$, requires a small mass $m_\phi \sim 10^{-5}~ \mpl$. In order to prevent quantum corrections from renormalizing $m$ up to $\mpl$, the inflation must possess tiny couplings to itself and all SM fields~\cite{nimatalk}. While small couplings protect $m \ll \mpl$, they do not present a satisfactory explanation for \emph{why} $m_\phi$ is so much smaller than $15 ~\mpl$, which is the VEV $\phi$ must traverse in $m^2 \phi^2$ inflation. One might relegate this problem to an initial conditions puzzle~\cite{Linde:1983gd}, though a deep understanding of what a super-Planckian $\Delta \phi$ implies remains elusive~\cite{Hawking:1987bi,Gibbons:1986xk,Kofman:2002cj,Hollands:2002xi,Gibbons:2006pa,Carroll:2010aj,Downes:2012xb,Remmen:2013eja}.

Similar theoretical complaints can be levied against Planck flat inflation. In a standard realization of Planck flat inflation, inflection point inflation \cite{Lyth:1998xn,Kachru:2003sx,Kallosh:2004yh,BlancoPillado:2004ns,Allahverdi:2006iq,Linde:2007jn,Chen:2009nk,Mazumdar:2010sa,Allahverdi:2011su,Agarwal:2011wm,Downes:2011gi,Downes:2012xb}, there needs to be tuning between the couplings of the cubic and linear terms of the potential $V = A \phi^3 + B \phi$ in order for the potential to be sufficiently flat to generate the required number of e-folds. For a $\phi^3$ potential like the one shown, the typical tuning required between $A$ and $B$ is of order $10^3$ \cite{Linde:2007jn,Downes:2011gi}.

Putting aside tuning arguments, it can be easily demonstrated that for a primordial tensor to scalar perturbation ratio $r \gtrsim 0.01$, and  for \textit{monotonically increasing} $\epsilon_V$, single field slow-roll inflation must be Planck fat. The exponential expansion of a single slowly rolling scalar field is related to its potential energy through the ratio of its tensor and scalar cosmological perturbations. Starting with the equations of motion and continuity for single field inflation,
\begin{align}
\ddot{\phi} + 3 H \dot{\phi} + V_\phi  &= 0, \label{eq:eom}\\
\frac{1}{3 \mpl^2} \left(V + \frac{1}{2} \dot{\phi}^2 \right)  &= H^2 \label{eq:cont},
\end{align}
and their slow-roll approximated versions, $3 H \dot{\phi} =- V_\phi$ and $H^2 = \frac{V}{3 \mpl^2} $, we recast the canonical e-fold integral in terms of $\phi$,
\begin{align}
N = \int_{t_*}^{t_e} H {\rm d}t = \mpl^2 \int_{\phi_e}^{\phi_*} \frac{V}{V_\phi} {\rm d}\phi. \label{eq:e-folds}
\end{align}
Varying the Einstein-Hilbert action with respect to scalar and tensor degrees of freedom leads to expressions for the dimensionless primordial scalar and tensor power spectra,
\begin{align}
A_s &= \frac{V}{24 \pi^2 \mpl^4 \epsilon_V}, \label{eq:canonicalAs}\\
A_t &= \frac{2V}{3 \pi^2 \mpl^4}. \label{eq:canonicalAt}
\end{align}
Defining $r \equiv A_t/A_s$ as the ratio of tensor to scalar modes and using the definition of $\epsilon_V$ along with the formulae for the scalar and tensor spectra, Eqs.~(\ref{eq:canonicalAs}) and (\ref{eq:canonicalAt}), we can recast Eq.~(\ref{eq:e-folds}) as
\begin{equation}
\frac{\Delta \phi}{\mpl} \gtrsim N \sqrt{\frac{r}{8}} = N \sqrt{2 \epsilon_V},
\label{eq:lythb}
\end{equation}
where we assume $\epsilon_V$ is monotonic increasing. Eq.~(\ref{eq:lythb}) is known as the Lyth Bound \cite{Lyth:1996im}\footnote{The small field bound of \cite{Lyth:1996im} assumes a scale-invariant power spectrum. The constraint presented in this paper gives a quantitative bound for a non-zero spectral index.} or Boubekeur-Lyth Bound \cite{Boubekeur:2005zm}. For inflationary scenarios which use a single slowly rolling scalar, large $r$ ($\gtrsim 0.01$) implies large $\epsilon_V$. Assuming that $\epsilon_V$ increases monotonically (until it equals unity and inflation ends) and combining the value of $\epsilon_V$ with $N \sim 60$, the Lyth Bound indicates that super-Planckian field excursions are necessary. 

\begin{figure}
\includegraphics[scale=.35]{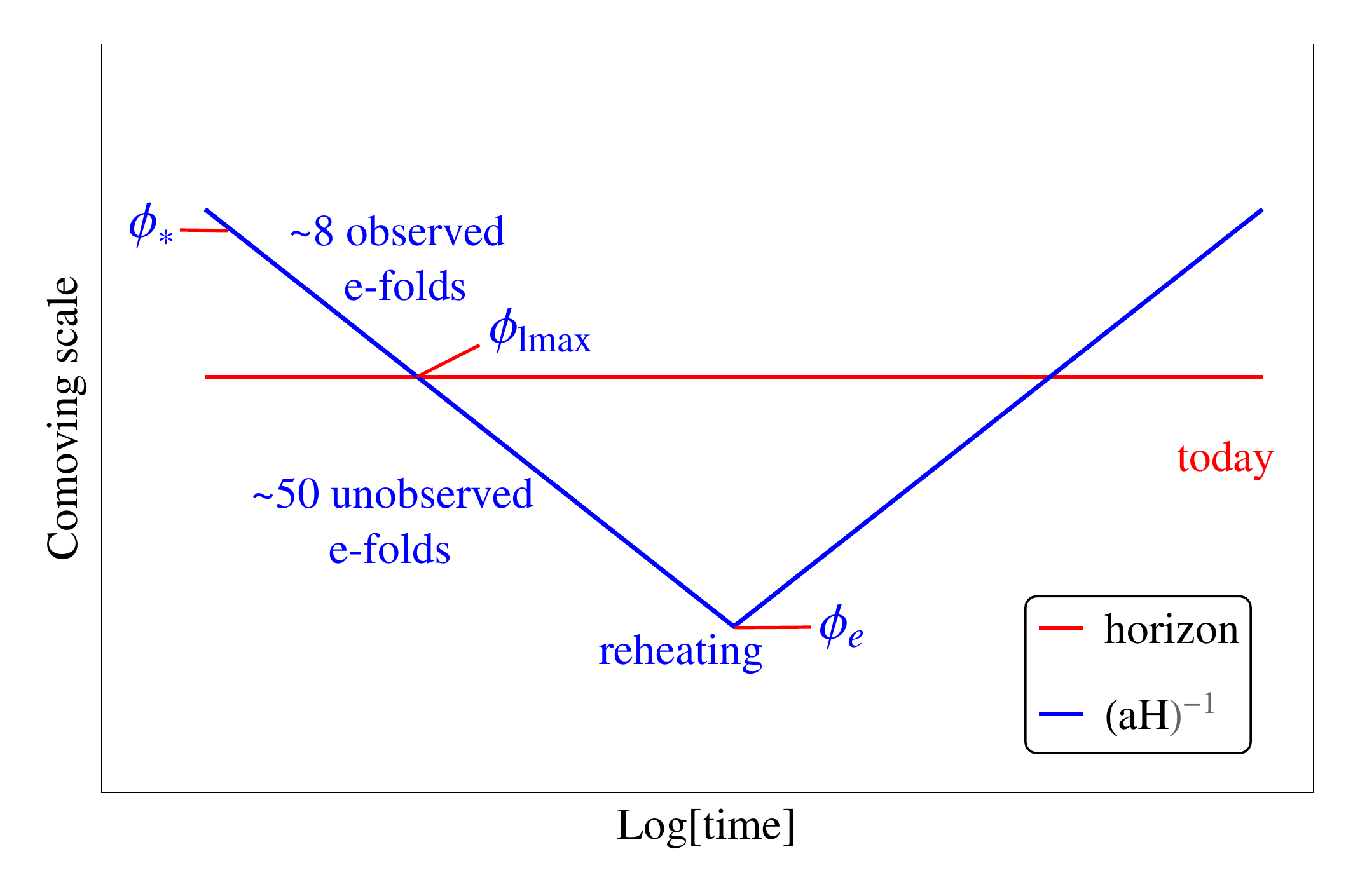}
\caption{This illustration shows that the smaller, unobserved primordial scales exited the de Sitter horizon after the larger modes observed today in the CMB. We have also indicated points on the comoving horizon where modes are sourced by $\phi$ at the pivot scale $\phi_*$, and also the point beyond which we have not made perturbation measurements, $\phi_{\ell \rm  max}$, and finally the point at which inflation ends, $\phi_e$.}
\label{fig:conform}
\end{figure}

\vskip1ex
However, there are steps one can take to evade or loosen the Lyth Bound on single field inflation. The first step is to recall that we have only observed a fraction of the primordial scalar power spectrum \cite{Ackerman:2010he}. The CMB only captures the first eight of the sixty observationally required inflationary e-foldings, which are followed by another $\sim 50$ e-folds we have not observed. In Figure \ref{fig:conform} we plot the classic, didactic picture showing the evolution of the comoving physical scale during inflation. The smallest scale at which primordial scalar perturbations have been measured is $k_{\ell \rm  max} \sim 0.2 ~{\rm Mpc^{-1}}$ \cite{Ade:2013kta,Ade:2013zuv,Ade:2013uln}. The total number of observed e-folds implied by this smallest scale is $N = {\rm log (0.2 ~{\rm Mpc^{-1}}/H_0)\sim 8}$. 

The second step is to loosen the requirements on $\eta_V$ and $\epsilon_V$. The Lyth bound assumes both $\eta_V$ and $\epsilon_V$ are small and that $\epsilon_V$ is monotonic increasing, neither of which is necessary. As examined in the modified slow-roll framework of \cite{Dvorkin:2009ne}, $\eta_V$ can momentarily reach very large values without disrupting inflation. 

\begin{figure}
\includegraphics[scale=.35]{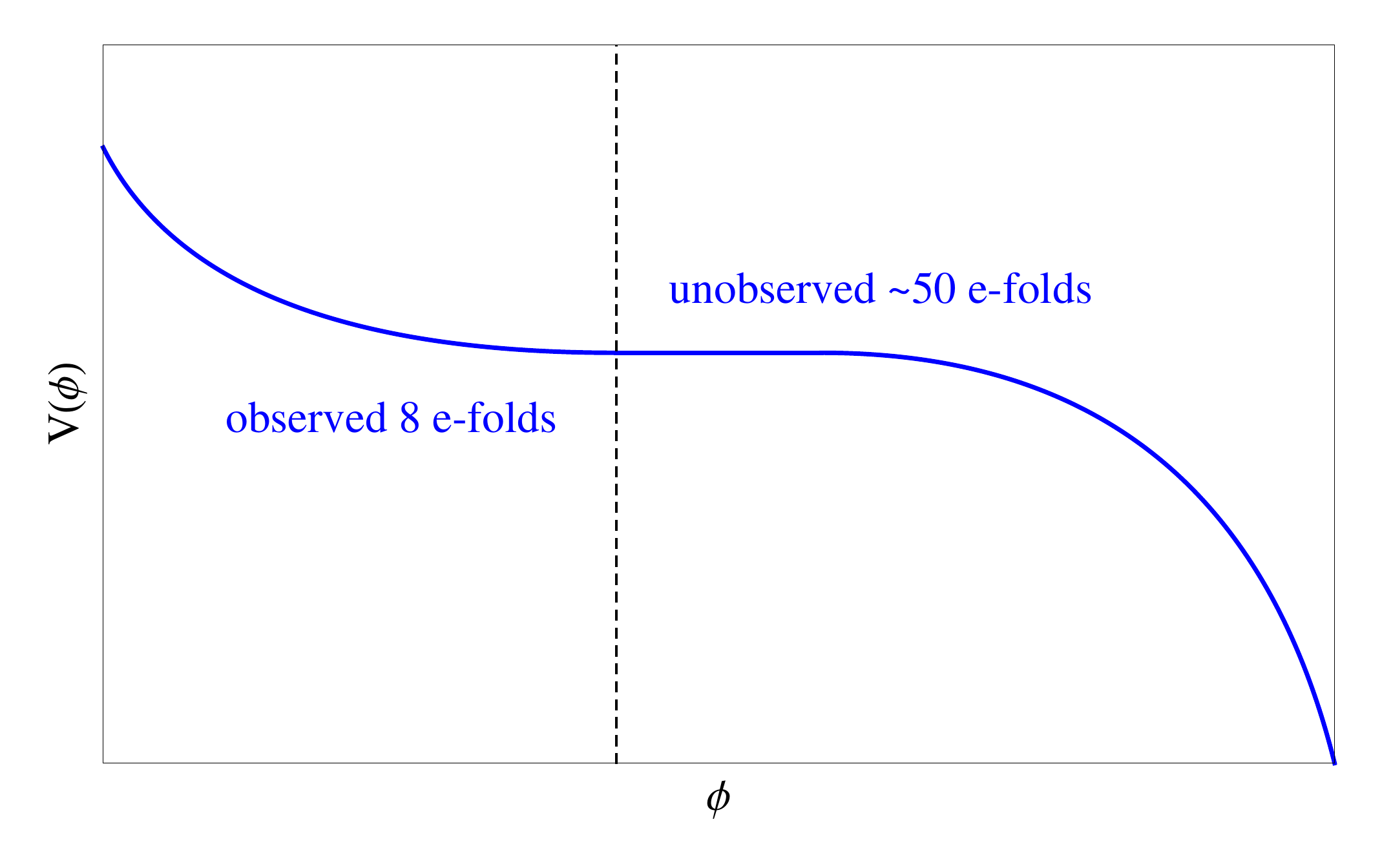}
\caption{The shape of an inflationary potential is depicted, which begins with $\epsilon_V$ monotonic increasing, before abruptly becoming flat. The flat part of the potential implies that the slow roll parameters take very small values ($\epsilon_V \ll 1$). This in turn implies an increase in the scalar power spectrum for modes outside observable (CMB) primordial scales.}
\label{fig:lythevader}
\end{figure}

Taking both of these points into consideration, one is lead to two-epoch potentials of the form shown in Fig. \ref{fig:lythevader}. The potential initially has large, monotonically increasing $\epsilon_V$ (first epoch), but transitions to a nearly flat regime causing $\epsilon_V$ to plummet to small values (second epoch). Potentials of this shape are similar to those considered in \cite{BenDayan:2009kv,Hotchkiss:2011gz}. The large $\epsilon_V$ in the initial epoch generates the observed $r \gtrsim 0.05$ and redshifted power spectrum $n_s - 1 \simeq -0.04$, while the second epoch of small $\epsilon_V$ provides the bulk of the required 60 e-folds. Between the two epochs there is a transition region where $\eta_V$ and $\epsilon_V$ parameters vary significantly, thereby invalidating the Lyth Bound (when taken over all 60 e-folds). Such behavior in $\epsilon_V, \eta_V$ is feasible so long as these variations, which imply observable variations in the power spectrum, hide within the latter 50 e-folds. Since $\epsilon_V$ is large only for a fraction of the trajectory in these setups, $\Delta \phi < \mpl$ is possible, and single-field, sub-Plankian models can still be viable. The main goal of this paper is to impose further constraints on setups which, like these two-epoch, single field inflation scenarios, vary $\epsilon_V$ and $\eta_V$. In this study we will consider a range of $r$ values: $0.05 \lesssim r \lesssim 0.2$. Our calculations are fairly general, and the resulting bounds are relevant for any value of $r \gtrsim 0.01$ measured by BICEP2 or other pending experiments.

We find constraints on both the observed and unobserved epochs. For the first epoch, conventional inflaton models that match CMB observations struggle to remain sub-Planckian even when we require they only generate the $\sim 8$ e-folds of observed inflation. Thus, with $\Delta \phi$ already nearing $\mpl$, there is little room in field space for the inflation to decelerate into the second epoch. To further complicate things, the slight redshift in the observed scalar power spectrum ($n_s-1=-0.04$) suggests that the inflaton was accelerating down the potential as the observed CMB modes left the horizon. A realistic sub-Planckian trajectory must not only cause this accelerating inflaton to decelerate just beyond the primordial scale reach of current experiments, but it must do so by a very large amount ($\epsilon_V \ll 1$) and rapidly.  
For the second epoch, even assuming the inflaton decelerates rapidly, the resulting tiny $\epsilon_V$ poses an interesting problem. If the inflaton moves too slowly ($\epsilon_V \ll 1$), the scalar power spectrum can become so large that it generate primordial black holes. These primordial black holes will be overabundant if the primordial power spectrum is too large for the smallest, as yet unmeasured, primordial modes.

The structure of our study is as follows: In Section \ref{sec:an} we reexamine the lower bound recently set by Antusch and Nolde (AN)~\cite{Antusch:2014cpa} on the super-Planckian excursion of a single slowly rolling inflaton, and find that it can be technically evaded if $|\eta_V| > 1$ (see Dvorkin and Hu \cite{Dvorkin:2009ne}). In Section \ref{sec:phenobound} we derive a phenomenological bound on small field inflation, which holds for $|\eta_V| > 1$ and non-monotonic $\epsilon_V$, by considering the Planck measurement $n_s-1 = -0.04$ over the 8 observed e-folds of inflation. In Section \ref{sec:8e-fold} we demonstrate explicitly that simple single field potentials normally considered will be nearly or actually super-Planckian for 8 e-folds of inflation, when matched with cosmological observables. Finally, in Section \ref{sec:pbh} we show that even if one wished to entertain a $\Delta \phi > 0.7 ~ \mpl$ trajectory as sub-Planckian, avoiding the formation of an over-abundance of black holes requires $\epsilon_V$ to become rapidly small (but not too small), then remain small and constant for 40-50 e-folds, then speed up rapidly to end inflation. We conclude in Section \ref{sec:conc}.


\section{Understanding the Antusch and Nolde Bound}
\label{sec:an}

Recently, Ref.~\cite{Choudhury:2014kma} claimed to have found a model of $\Delta \phi < 0.1 ~\mpl$, single field inflation consistent with a scalar-to-tensor ratio of $r \geq 0.1$ and all Planck observations, derived by fitting slow-roll parameters of the field's trajectory to fourth order in $\dot{H}$. In response to this paper, Antusch and Nolde (AN) \cite{Antusch:2014cpa} found a relationship between the number of e-folds achievable in a single field, slow-roll model, and the average of the difference between $\eta_V$ and $2\epsilon_V$ over the field range of $\Delta \phi$,
\begin{align}
\frac{\Delta \phi}{\mpl} \gtrsim \frac{0.11}{\left\langle \eta_V - 2 \epsilon_V \right\rangle_{\Delta \phi_{\rm min} 60}} \sqrt{\frac{r}{0.1}},
\label{eq:an}
\end{align}
where ``$\left\langle ~ \right\rangle_{\Delta \phi_{\rm min}60}$" indicates the average of this term over the field range $\Delta \phi_{\rm min} = \phi_* - \phi_{\rm min}$ of a single, slowly rolling inflaton sourcing 60 e-folds. We define $\phi_{\rm min}$ as the value of $\phi$ where $\epsilon_V$ is at its minimum along a trajectory. Under the assumption that $|\eta_V| \ll 1$ and $\epsilon_V \ll 1$, this implies that even if $\epsilon_V$ is a varying function, it is not possible to achieve sub-Planckian inflation with large values of $r$ over 60 e-folds. The authors of the 4th order slow-roll parameter model responded with a spirited defense, Ref.~\textcolor{red}{ \cite{Choudhury:2014wsa}}.

In this paper we will not address the fitted fourth order slow-roll model detailed in \cite{Choudhury:2014kma} and \textcolor{red}{ \cite{Choudhury:2014wsa}}. We will, however, examine the bound of Antusch and Nolde Eq.~(\ref{eq:an}), in light of the fact that $|\eta_V| \gg 1$ is permitted over a few e-folds in a general slow-roll framework \cite{Dvorkin:2009ne}. 

First we review the AN Bound. The derivative of $\sqrt{2 \epsilon_V}$ with respect to the inflaton, $\phi$, is
\begin{align}
\mpl \frac{d}{d \phi} \sqrt{2 \epsilon_V} = \mpl^2 \left[ \frac{V_{\phi \phi}}{V} - \frac{V_\phi^2}{V^2} \right] = \eta_V - 2 \epsilon_V.
\end{align}
This is integrated over $\phi$, from a field value where $\epsilon_V$ is at its minimum, $\phi_{\rm min}$, to the field value at the start of Hubble scale inflation, $\phi_*$,
\begin{align}
\mpl^3 \left[\left( \frac{V_\phi}{V} \right)_* - \left( \frac{V_\phi}{V} \right)_{\rm min}\right] = \mpl \int_{\phi_{\rm min}}^{\phi_*} (\eta_V - 2\epsilon_V){\rm d} \phi  \nonumber \\ = \mpl (\phi_* - \phi_{\rm min}) \left\langle \eta_V - 2\epsilon_V \right\rangle_{\Delta \phi_{\rm min} 60}. \label{eq:longveq}
\end{align}
The final equality is a definition: the average value of the integrand ``$\left\langle \right\rangle_{\Delta \phi_{\rm min} 60}$" multiplied by the range of the integral is equivalent to the integral. This quantity is important for interpreting the AN Bound. 

It remains to show that the term $(V_\phi/V)_{\rm min}$ may be discarded because it is small compared to $(V_\phi/V)_*$. Assuming $r \gtrsim 0.1$, and combining Eq.~(\ref{eq:canonicalAs}) and Eq.~(\ref{eq:canonicalAt}) into $r = 16 \epsilon_V$, one can easily see that
\begin{align}
\mpl \left( \frac{V_\phi}{V} \right)_* = \sqrt{2 \epsilon_V} = 0.14 \sqrt{\frac{r}{0.16}}. \label{eq:anbig}
\end{align}
On the other hand, we can set a ceiling on the value of $(V_\phi/V)_{\rm min}$,
\begin{align}
\phi_* - \phi_e = \int_{N_e}^{N_*} \frac{{\rm d} \phi}{{\rm d} N}{\rm d} N = \int_{N_e}^{N_*} \frac{{\rm d} \phi}{H {\rm d} t}{\rm d} N = -\int_{N_e}^{N_*} \frac{V_\phi}{3 H^2}{\rm d} N \label{eq:ansmall1} 
\\ |\Delta \phi| = \mpl^2 \int_{N_e}^{N_*} \frac{V_\phi}{V}{\rm d} N > \mpl^2 N_{\rm total} \left( \frac{V_\phi}{V} \right)_{\rm min}, \label{eq:ansmall}
\end{align}
where $N_{\rm total}$ is the total number of e-folds. In Eq.~(\ref{eq:ansmall1}) we have used the slow-roll equations of motion Eq.~(\ref{eq:eom}) and continuity Eq.~(\ref{eq:cont}) in the third and fourth equalities. In Eq.~(\ref{eq:ansmall}) we use the fact that the last term is smaller than the integral it follows after, because any integral is larger than the smallest value of the integrand multiplied by the range of the integral.\footnote{There are some subtleties to Eq.~(\ref{eq:ansmall}), because if the smallest value of $(V_\phi/V)_{\rm min}$ is negative, but $(V_\phi/V)$ oscillates around zero over the $\phi$ range of the integral, then it no longer follows that $|\phi_* - \phi_e| = \Delta \phi > |\mpl^2 N_{\rm total} \left( V_\phi/V \right)_{\rm min}|$, which is required for the AN bound. An oscillation of $(V_\phi/V)$ implies a model with oscillating $\epsilon_V$. One example of an oscillating single field potential is resonant non-Gaussianity \cite{Flauger:2010ja}. We can probably exclude most single field models with large oscillating $\epsilon_V$ over the observed 8 e-folds, because this would imply oscillations of the scalar power spectrum.} Then Eq.~(\ref{eq:ansmall}) implies that $(V_\phi/V)_{\rm min} < (1/60) (\Delta \phi / \mpl)$, where we assume 60 e-folds of inflation. Comparing this to Eq.~(\ref{eq:anbig}) we see that $(V_\phi/V)_{\rm min}$ can be neglected for the purposes of this bound. 

With the final observation that $\Delta \phi \geq \Delta \phi_{\rm min} = |\phi_* - \phi_{\rm min}|$ and combining Eq.~(\ref{eq:anbig}) and Eq.~(\ref{eq:longveq}), we find that
\begin{align}
\frac{\Delta \phi}{\mpl} \gtrsim \frac{0.14}{\left\langle \eta_V - 2\epsilon_V \right\rangle_{\Delta \phi_{\rm min 60}}} \sqrt{\frac{r}{0.16}}.
\label{eq:anbound}
\end{align}
It is clear that for small values of $\eta_V$ and $\epsilon_V$, any $r$ greater than $0.1$ requires a super-Planckian field trajectory.

However, it is permissible for $|\eta_V|$ be of order unity over a number of e-foldings. Indeed, for the very same Planck fat to Planck flat inflationary trajectory discussed in Section \ref{sec:intro}, this fat to flat transition occurs at the point in the inflaton trajectory where $|\eta_V|$ becomes order unity and $\epsilon_V$ becomes tiny while the inflaton sources primordial modes smaller than those that have been observed. This behavior can be seen in Figure \ref{fig:lythevader}. The inflaton can come to near rest ($\epsilon_V \ll 1$) rather abruptly, which requires $|\eta_V|\gtrsim1$ over some amount of the field trajectory. The question then becomes whether the width of the trajectory in e-folds multiplied by the average large value of $\left\langle \eta_V - 2\epsilon_V \right\rangle$ in this region is greater than $\sim \mathcal{O}(1)$. If it is, the Antusch and Nolde bound as stated does not necessarily restrict these models. For example if $\eta_V - 2\epsilon_V$ is zero over 50 e-folds, but $|\eta_V|>10$ over 10 e-folds, this inflaton technically evades the bound of Antusch and Nolde.

In Appendix \ref{app:inflec}, we give an explicit example of an inflection point trajectory that exhibits $\left\langle \eta_V - 2\epsilon_V \right\rangle \gtrsim 1$, a result at odds with the AN Bound as stated. In fact, this same inflection point behavior examined in Appendix \ref{app:inflec} also arises in polynomial fit potentials in the literature \cite{BenDayan:2009kv,Hotchkiss:2011gz} which match CMB observables, traverse barely sub-Planckian field ranges, and have $r \sim 0.1$. These potentials become rapidly flat when sourcing e-folds over the smallest scales -- in doing so their trajectories have $|\eta_V| \gg 1$ over a significant fraction of e-folds, which is why the AN bound Eq.~(\ref{eq:anbound}) does not apply.

\section{A Phenomenological Bound on Small Field Inflation}
\label{sec:phenobound}

The AN bound is more general than the Lyth Bound in that it allows for varying $\epsilon_V$, however it is possible to set a stronger constraint. As we will now show, by using some relations in the AN derivation we can set a strong bound on small field inflation without requiring $|\eta_V| < 1$. The bound in this paper requires only the known measurement of $n_s-1$ over the observed 8 CMB e-folds \cite{Ade:2013kta,Ade:2013uln,Ade:2014xna} and holds for any theory of inflation with a spectral index sourced by first and second order slow-roll parameters. The bound hinges on the fact that while for a generic theory, $\eta_V - 2\epsilon_V$ may vary, for a red spectral index this quantity must average to less than one-third the measured value of $n_s-1$ over the observed 8 e-folds of primordial scalar modes. There is prior work bounding the field range of a single inflaton by incorporating the spectral index of primordial scalar perturbations \cite{Easther:2006qu}, although the form and interpretation of the bound in this work differs from that of Ref.~\cite{Easther:2006qu}.

We begin our derivation by reiterating Eq.~(\ref{eq:longveq}), and noting that for this bound we will not discard $\left(V_\phi/V \right)_{\rm min}$,
\begin{align}
\mpl^2 \left[\left( \frac{V_\phi}{V} \right)_* - \left( \frac{V_\phi}{V} \right)_{\rm min}\right]~~~~~~~~~~~~~~~ \nonumber \\= (\phi_* - \phi_{\rm min}) \left\langle \eta_V - 2\epsilon_V \right\rangle_{\Delta \phi_{\rm min}8}, \label{eq:newlongveq}
\end{align}
where we note that for this bound, we will only consider the observed 8 e-folds of inflation, so $\left\langle \eta_V - 2\epsilon_V \right\rangle_{\Delta \phi_{\rm min}8}$ is the average value of $\eta_V - 2\epsilon_V$ over the observed 8 e-folds of the CMB.

From Eq.~(\ref{eq:ansmall}) we see that $ \frac{\Delta \phi}{N_{\rm total}} \geq \mpl^2 \left( \frac{V_\phi}{V} \right)_{\rm min}  $ and $\mpl \left( \frac{V_\phi}{V} \right)_* = \sqrt{\frac{r}{8}}$. Hence again identifying $\Delta \phi \geq \Delta \phi_{\rm min} = |\phi_* - \phi_{\rm min}|$, we find that
\begin{align}
\frac{\Delta \phi}{\mpl}\geq \frac{\sqrt{\frac{r}{8}}}{\left\langle \eta_V - 2\epsilon_V \right\rangle_{\Delta \phi_{\rm min}8} + \frac{1}{N_{\rm total}}}. \label{eq:newlongveq2}
\end{align}
It remains to evaluate $\left\langle \eta_V - 2\epsilon_V \right\rangle_{\Delta \phi_{\rm min}8}$. To second order in slow-roll parameters, the spectral index is $n_s-1 = 2 \eta_V - 6 \epsilon_V$. The Planck collaboration has measured the spectral index at $n_s -1 = -0.04 \pm 0.0073$ \cite{Ade:2013kta,Ade:2013uln,Ade:2013zuv}, using modes from 8 e-folds of inflation observed in the CMB. Of course, $ \eta_V - 2\epsilon_V = \frac{1}{2}(n_s -1) + \epsilon_V$. For the argument that follows, the most conservative bound is attained if we assume $\epsilon_V$ contributes a dominant portion of the average value of the spectral index over the observed 8 e-folds. Because we know $n_s-1$ is negative and will make the denominator of Eq.~(\ref{eq:newlongveq2}) smaller, this means setting $\epsilon_V = - \frac{1}{6}(n_s -1)$, which leads to
\begin{align}
\left\langle \eta_V - 2\epsilon_V \right\rangle_{\Delta \phi_{\rm min}8} \leq \frac{1}{3}\left(n_s-1 \right) \label{eq:ns}
\end{align}
where both terms of this inequality are negative.

Placing this inequality in Eq.~(\ref{eq:newlongveq2}), it follows that for a negative spectral index,
\begin{align}
\frac{\Delta \phi}{\mpl}\geq \frac{\sqrt{\frac{r}{8}}}{\frac{1}{3}(n_s-1) + \frac{1}{N_{\rm total}}}. \label{eq:theboundanalytic}
\end{align}
Note that Eq.~(\ref{eq:theboundanalytic}) yields the Lyth Bound in the limit $n_s-1 \rightarrow 0$. Inserting the measured 8 e-folds and the spectral index into Eq.~(\ref{eq:theboundanalytic}), the bound on small field inflation is
\begin{align}
\frac{\Delta \phi}{\mpl} \geq 1.00 \sqrt{\frac{r}{0.1}}. \label{eq:thebound}
\end{align}
This bound relies only on inflation occurring over the measured 8 e-folds, and not on any particular behavior of $\epsilon_V$ or $\eta_V$. One might wonder how varying the Planck measurement of $n_s-1$ affects this bound. If we loosen the bound by assuming a $5\,\sigma$ deviation from the measured value of $n_s -1$, the bound barely weakens, $\frac{\Delta \phi}{\mpl} \geq 0.90 \sqrt{\frac{r}{0.1}}$.

This bound has interesting implications. It requires only the measured value of the spectral index, and that the spectral index be mainly determined by first and second order slow-roll parameters. Insofar as the Planck and WMAP measurements can be trusted, and we wish to construct a theory which does not require more slow-roll parameters than are required to fit the Planck data \citep{Ade:2013kta}, $r \gtrsim 0.05$ indicates $\Delta \phi \gtrsim 0.71 ~ \mpl$. Although one may exist, we are not aware of any examples of a single field model fitting the CMB power with a spectral index whose average value receives predominant contributions from third order or higher order slow-roll parameters. A large higher order slow-roll parameter would imply large shifts in the power spectrum which are not observed on the CMB \cite{Ade:2013kta,Ade:2013uln,Ade:2014xna}.

Future proposals to measure the scalar power spectrum using $\mu$- and $i$-type distortions of the CMB blackbody spectrum forecast sensitivity to 17 e-folds~\cite{Hu:1992dc,Khatri:2012tw,Sunyaev:2013aoa,Khatri:2013dha}. Assuming the scalar power spectral index on these scales stays near $n_s-1 = -0.04$, this will strengthen the small field bound of this paper to
\begin{align}
\frac{\Delta \phi}{\mpl} \geq 2.46 \sqrt{\frac{r}{0.1}}. \label{eq:themuibound}
\end{align}

The bound of this section indicates that it should be impossible to take the simplest inflationary potentials, calculate their observables and match them to Planck data, and find $\Delta \phi \ll \mpl$ for $r \gtrsim 0.1$, which we will confirm in the next section.

\section{Eight e-folds: Simple Potentials are Super-Planckian if $r\gtrsim0.1$}
\label{sec:8e-fold}

In prior sections we showed that for $r \gtrsim 0.05$, single field, slowly rolling inflationary potentials can be technically sub-Planckian if $\epsilon_V$ evolves non-monotonically and if $|\eta_V|$ becomes large or if the slow-roll parameters oscillate. However, the requirement that the spectral index $n_s-1 = 2 \eta_V - 6 \epsilon_V = -0.04$ over 8 e-folds and the bound given in Eq.~(\ref{eq:thebound}) indicated that any such models would have nearly Planckian field excursions anyway. In this section we demonstrate this explicitly, by examining some simple scalar polynomial potentials and requiring that they match observed cosmological parameters for 8 e-folds of inflation. 

As expected, we will not find any nice examples of sub-Planckian trajectories which fit the observed 8 e-folds of cosmological perturbations and which can be easily patched together with an inflection point to remain sub-Planckian over the last 50 e-folds to form a very sub-Planckian trajectory, like the one sketched in Figure \ref{fig:lythevader}. In other words, when we examine typical inflationary potentials and require they remain within $2 \sigma$ of WMAP and Planck measurements of the CMB, also requiring $r \sim 0.1-0.2$, although a few will be marginally sub-Planckian, they are anyway very nearly Planckian and so do not provide a simple theory which would be perturbative under order one corrections at a Planck cutoff.

\subsection{Quadratic potential}
For example, consider the potential $V(\phi) = m^2 \phi^2$. For this potential, $\epsilon_V = 2 \mpl^2 /\phi_*^2$. Since to first order in slow-roll parameters,\footnote{We note that all calculations in this section will be done to first order in the slow-roll approximation.} $r = 16 \epsilon_V$, we can rearrange $\epsilon_V = 2 \mpl^2 /\phi_*^2$ to find $\phi_* = 4 \mpl \sqrt{2/r}$. Then we calculate the number of e-folds associated with this $m^2 \phi^2$ potential (actually, up to a constant out front, the following expression will be generic for any potential polynomial in $\phi$), 
\begin{align}
	N_* = 8 = \frac{1}{\mpl^2}\int_{\phi_e}^{\phi_*} \text{d}\phi \frac{V}{V_\phi} &=
	\frac{1}{\mpl^2} \int_{\phi_e}^{\phi_*}\text{d}\phi \frac{\phi}{2} \\ &= \frac{1}{4 \mpl^2} \left(
	\phi_*^2 - \phi_e^2 \right).
	\label{}
\end{align}
Inserting $\phi_* = 4 \mpl \sqrt{2/r}$ and solving for $\phi_e$ gives
\begin{equation}
	\phi_e = \mpl \sqrt{32\left( \frac{1}{r} - 1 \right)} .
	\label{}
\end{equation}
Combining these formulae and enforcing $r=0.1$, we find that $\Delta \phi = (17.89-16.97) ~\mpl = 0.92 ~ \mpl$.\footnote{It is trivial to rescale the scalar field in an $m^2 \phi^2$ potential so that the field range shown here as $\Delta \phi = (17.89-16.97) ~\mpl= 0.92 ~\mpl$ terminates at a field value of zero. The resulting potential will remain polynomial in $\phi$.} However, having fit $r$ and $N$ with $\phi_*$ and $\phi_e$, we have no free parameters remaining to fit the spectral index, which at $\phi_*$ is $n_s-1 = -0.025$, more than $2 \sigma$ away from the Planck result. Increasing $r$ to $0.2$ results in $\Delta \phi = (12.65 - 11.31) ~\mpl= 1.34~ \mpl $. Regardless, we see that to match $m^2 \phi^2$ to the observed 8 e-folds of inflation, if we also require $r=0.1$ and that the field range remain sub-Planckian, we do not have enough free parameters to comfortably match the value of $n_s$ measured by Planck.

\subsection{Power law potential}
Finding that $m^2 \phi^2$ fails to reproduce CMB observables over 8 e-folds while remaining sub-Planckian, we can ask whether any single polynomial term can reproduce the primordial scalar perturbations observed, with $r \gtrsim 0.1$. We will see that allowing $n$ to vary in a general power-law potential of the form $V(\phi) =\lambda \phi^n$, where the energy dimension of $\lambda$ is given by $[\lambda] = 4-n$, barely allows both $n_s$ and $r$ to be fit for a polynomial potential sourcing 8 e-folds. For a generic power law potential, $\epsilon_V = n^2 \mpl^2 / (2\phi_*^2)$, so
\begin{equation}
	r = \frac{8 n^2 \mpl^2}{\phi_*^2} .
	\label{}
\end{equation}
Finding the spectral index to first order in $\epsilon_V$ and $\eta_V$ gives
\begin{equation}
	n_s = 1 + 2\eta_V - 6\epsilon_V = \frac{\phi_*^2 - n (n+2)\mpl^2}{\phi_*^2} .
	\label{}
\end{equation}
Since there are 2 observables, $n_s$ and $r$, we can solve for both $n$ and $\phi_*$ (using $n_s = 0.96$). Then relating the number of e-folds to $\phi_e$ gives
\begin{equation}
	N_* = 8 = \frac{1}{\mpl^2}\int_{\phi_e}^{\phi_*} \text{d}\phi \frac{\phi}{n} =
	\frac{1}{2n \mpl^2} \left( \phi_*^2 - \phi_e^2 \right).
	\label{}
\end{equation}
From this it follows that
\begin{equation}
	\phi_e = \sqrt{\phi_*^2 - 16 n \mpl^2 } .
	\label{}
\end{equation}
For $r=0.1$, $n=0.919$ and $\Delta \phi = 8.22-7.27 ~\mpl= 0.95 ~\mpl$. This result is at first surprising, because it is at odds with the aforestated bound of Eq.~\ref{eq:thebound}. However, we note that we have found a model where $n_s-1=-0.04$ at $\phi_*$. At $\phi_e$, $n_s-1\sim-0.03$ -- which makes the bound given in its full form in Eq.~\ref{eq:newlongveq2} weaker, because $n_s-1$ no longer averages to $-0.04$ over the field range. Nevertheless, this model falls within $\sim 2 \sigma$ of the Planck result. Hence we see that the simplest polynomial potential for $r=0.1$, which barely fits bounds on the spectral index and power spectrum over 8 e-folds, is nearly linear in $\phi$.\footnote{For $r=0.2$, the viability of a simple sub-Planckian polynomial inflation no longer holds at all. For $r=0.2$, $n=3.40$ and $\Delta \phi = 21.51 - 20.20 = 1.31 $.} Finally, we note that for this type of potential, the scalar perturbation amplitude $A_s$ can be fit with the parameter $\lambda$.

A number of common inflationary potentials and their field range over 8 e-folds are included in Appendix \ref{app:pots}. The smallest field range obtained for a potential which fits Planck observations at $\phi_*$ is that of an exponential potential, where $\Delta \phi$ can be as ``small" as 0.89 $\mpl$ over 8 e-folds. Once again, the apparent departure from the bound given in Eq.~\ref{eq:thebound} is actually a result of the spectral index not averaging to $n_s-1=-0.04$. We see that generic single field inflationary potentials with $r \gtrsim 0.1$ are only sub-Planckian if $n_s$ is bluer than Planck indicates.

\section{Bounds on Small Field Inflation from Primordial Black Holes}
\label{sec:pbh}

We showed in earlier sections that, while the first period of inflation produces the perturbations we have observed in the CMB, a substantially different second inflation would be necessary in order to account for both a small inflaton field range and the total flatness and causal connection of the horizon we observe. It should be emphasized that the model building goal in such a setup would be to fit both epochs of inflation into a single field model which transits a sub-Planck field range. Despite the complexity implied, if we continue the quixotic task of patching together a model of barely sub-Planckian inflation which fits observation, we next need to find what constraints can be put on the unobserved 50 e-folds of inflation depicted in the latter half of Figure \ref{fig:lythevader}. 

Over these 50 e-folds, $\epsilon_V$ must become small for $\Delta \phi$ to remain sub-Planckian. Because $A_s \propto V(\phi)/\epsilon_V$, this second inflation epoch tends to have large power. Too much power at any given $k$ has been shown in Ref.~\cite{Josan:2009qn} to result in abundant underinflated patches of space that seed primordial black holes during reheating. The primordial black hole abundance is constrained across an immense range of scales via astrophysical surveys of gravitational lensing, black hole evaporation, or an overclosed universe~\cite{Carr:1974nx,Green:1997sz,Kribs:1999bs,Bugaev:2002yt,Carr:2005zd,Zaballa:2006kh,Josan:2009qn}.  In this section we will use these constraints to limit the maximum energy scale for a second inflationary period with a fixed sub-Planckian inflaton field range $\Delta \phi < \mpl$. 

To derive an energy-level bound, we will make the simplifying assumption that the slow-roll parameter $\epsilon_V$ is constant across the fifty, as yet unobserved e-folds of inflation. For constant $\epsilon_V$, we can then apply the Lyth Bound~\cite{Lyth:1996im} to the second epoch to relate the amount of $\Delta \phi$ traversed  in the second epoch to $\epsilon_V$,
\begin{align}
\frac{\Delta \phi}{\mpl} \gtrsim N \sqrt{\frac{r}{8}} = N \sqrt{2 \epsilon_V}.
\label{eq:delphisecond}
\end{align} 
Note that the $\Delta \phi$ calculated in this equation will be in \emph{addition} to the $\Delta \phi$ traversed during the observed 8 efolds of inflation. We recast (\ref{eq:delphisecond}) to find that for a constant $\epsilon_V$, $\epsilon_V = (\frac{\Delta \phi}{\sqrt{2} \mpl N})^2$. 

\begin{figure}
\centering
\includegraphics[scale=.5]{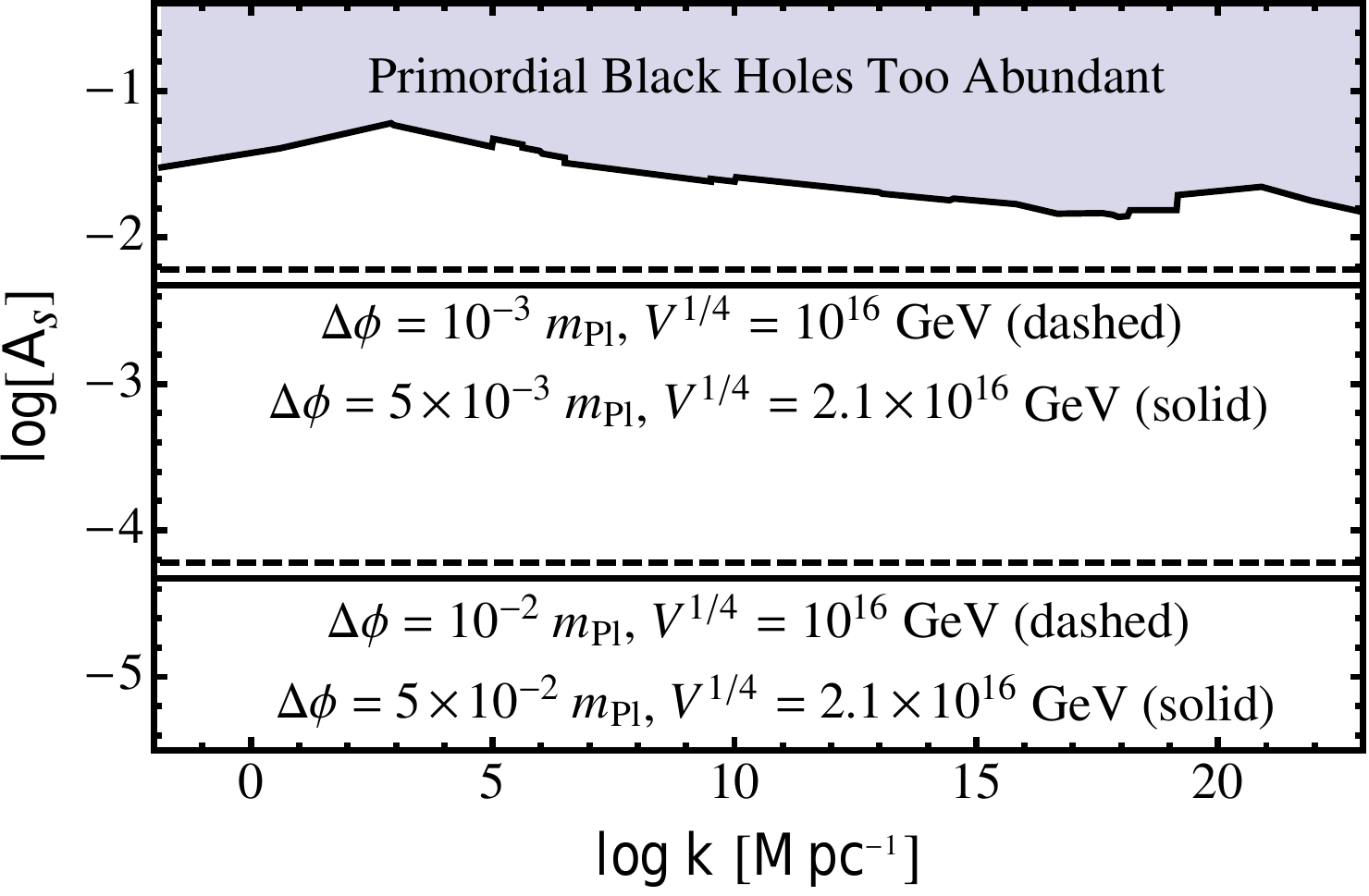}
\caption{The bound on total scalar curvature perturbations from primordial black hole formation is shown, which ranges from $k\sim 10^{-2} - 10^{23} ~ {\rm Mpc^{-1}} $, (adapted from \cite{Josan:2009qn}). Under the assumption that the slow-roll parameter $\epsilon_V$ remains fixed, the implied level of primordial scalar power is shown for 50 e-folds of inflation at energy scales $V^{1/4} = 2.1 \times 10^{16} ~{\rm GeV ~ and}~10^{16}$ GeV, for single field ranges $\Delta \phi \sim 10^{-2}-10^{-3} ~{\rm \mpl}$ as indicated.}
\label{pbhbound}
\end{figure}

The simplest bound we can derive is on $A_s$ itself, i.e. assuming that the second epoch of inflation is scale-invariant. Using Eq.~(\ref{eq:delphisecond}) and following~\cite{Carr:1974nx,Green:1997sz,Kribs:1999bs,Bugaev:2002yt,Carr:2005zd,Zaballa:2006kh,Josan:2009qn}, for a fixed sub-Planck inflaton range $\Delta \phi$ and a fixed energy scale of inflation $V$, black hole abundances limit the level of primordial scalar perturbations to be less than
\begin{align}
A_s = \frac{VN^2}{12 \pi^2 \mpl^4 \Delta \phi^2} \lesssim 10^{-2}.
\label{eq:asn}
\end{align}
In Figure \ref{pbhbound} we compare primordial black hole bounds on scalar power to the levels of scalar power implied by fixed values for $V$ and $\Delta \phi$ which permit a 50 e-fold sub-Planckian field excursion for a single field during the unobserved portion of inflation.

The recent BICEP2 measurement of $r \gtrsim 0.1$ implies that the scale of inflation is approximately $2.1 \pm 0.2 \times 10^{16}$ GeV  \cite{Ade:2014xna}. Taken at face value, Figure \ref{pbhbound} indicates that the 50 additional e-folds of inflation necessary to explain the structure of our universe can easily be accommodated in a sub-Planckian single field setup at the inflation energy scale found by BICEP2. Nevertheless, the treatment of scalar perturbations above can be improved. As we emphasized in our analysis of the AN Bound in Section \ref{sec:an}, in order to swiftly drive $\epsilon_V$ to small values one must assume the second slow-roll parameter $|\eta_V|$ is large. If $|\eta_V|$ is large, one must use a modified slow-roll formalism to calculate the power spectrum \cite{Dvorkin:2009ne}.

Such a detailed treatment of the power spectrum, which may exhibit oscillatory features during departures from $|\eta_V|\ll 1$ slow-roll \cite{Dvorkin:2009ne,Downes:2012xb}, requires choosing a specific potential and resorting to numerics. Since our goal is to be model independent, we instead estimate the field range required to transition from $A_s \sim 10^{-9}$ to $A_s \sim 10^{-2}$ for an inflaton trajectory like that shown in Figure \ref{fig:lythevader}.

Assuming that the trajectory of the inflaton is driving $\epsilon_V$ to very small values, $\epsilon_V$ can be neglected in determining the growth of the power spectrum. In this case, $n_s - 1$ is set by $\eta_V$ alone, which we assume is $\mathcal O(1)$ and constant. The scalar power spectrum during the transition from small to very small $\epsilon_V$ can then be approximated by
\begin{align}
A_s(k) = A_{0} \left(\frac{k}{k_{\ell {\rm max}}}\right)^{n_s-1} \simeq A_{0} \left(\frac{k}{k_{\ell {\rm max}}}\right)^{2 \eta_V},
\end{align}
where $k_{\ell {\rm max}} = 0.2 ~{\rm Mpc^{-1}}$ is the scale at which the CMB measurements end. From this we can easily find the number of e-folds over which the potential transitions to a very flat trajectory,
\begin{align}
{\rm log} \left(\frac{k}{k_{\ell {\rm max}}}\right) = N_{\rm trans} \simeq \frac{{\rm log} \left(\frac{A_s(k)}{A_{0}}\right)}{2 \eta_V} = \frac{16}{2 \eta_V},
\end{align}
where the last equality follows from assuming $A_s(k) \sim 10^{-2}$ and $A_{0} \sim 10^{-9}$. 

It would be useful to determine the field range $\Delta \phi$ that will be traversed during a fast transition, e.g. $\eta_V=1$. $\Delta \phi$ during this rapid diminution of $\epsilon_V$ can be estimated by taking the average of the slow-roll parameter at the scale $k_{\ell {\rm max}}$, $\epsilon_V^{(\ell \rm max)} \sim 0.023$, and its value after the transitional period, $\epsilon_V^{(f)} \sim 2.4 \times 10^{-9}$. These values of $\epsilon_V$ were both calculated with Eq.~(\ref{eq:asn}), assuming a fixed energy scale for inflation, $V = 2.1 \times 10^{16} ~{\rm GeV}$. Taking an intermediate value of $\epsilon_V \sim 10^{-5}$ and estimating the required field range to transition to $\epsilon_V^{(f)} \sim 2.4 \times 10^{-9}$ for $\eta_V =1$ (which amounts to calculating $\Delta \phi$ for $\epsilon_V\sim 10^{-5}$ over 8 e-folds), yields $\Delta \phi_{\rm trans} < 0.05 ~\mpl$. The same calculation done for $\eta_V > 1$ will of course result in an even smaller field range.

Combining this $\Delta \phi_{\rm trans}$ with the $\Delta \phi$ values from the first 8 e-folds, we see that it is technically possible to construct an inflaton trajectory which sources 50 e-folds over unobserved, small scales, with the inflaton remaining sub-Planckian in field range, and the power spectrum remaining below the maximum level set by primordial black hole bounds. Nonetheless, we note that the required trajectory will need to exhibit four distinct features: (1) $\epsilon_V$ and $\eta_V$ must begin at values consistent with Planck and WMAP measurements, (2) $|\eta_V|$ must become large to drive $\epsilon_V$ to small values, (3) $\eta_V$ must become nearly zero again to prevent $\epsilon_V$ from
becoming too small and violating primordial black hole constraints,\footnote{We note that, as $A_s$ is proportional to $V$, a second way to ameliorate excessive power is to arrange for $V$ in the second epoch to be much lower than in the first, i.e. that the second era of inflation occurs at a different, lower energy scale~\cite{Silk:1986vc,Holman:1991ht,Adams:1997de,Burgess:2005sb}. The potential could safely drop to this second energy scale after the observed GUT scale inflationary period, without upsetting current cosmological observations. We do not address this possibility at any length, because we have shown that assuming a rapid transition ($\eta_V \sim 1$) to $\epsilon_V\sim 10^{-9}$ is sufficient for the total field range to remain sub-Planckian.} and finally (4) $|\eta_V|$ must become large again, so that $\epsilon_V$ can become large and inflation can end rapidly. The fourth feature is necessary, for if on the contrary, inflation ends with $\epsilon_V$ slowly growing, even without considering the rest of the inflaton trajectory, this alone will imply a
Planckian field range for the inflaton.\footnote{We note that the behavior described here is particular to models attempting to fit high field behavior, a sub-Planckian field range, and all cosmological observations. This should not be confused with the behavior of simple inflection point models, for instance the one outlined in Appendix \ref{app:inflec}, which easily comes to an end after the inflaton rolls through the inflection point.}

\section{Conclusion}
\label{sec:conc}

For the first thirty years of primordial cosmology, a set of classical gravitational inconsistencies (chiefly horizon and flatness) led to the proposal of a new quantum phenomenon, the inflaton, with concomitant quantum predictions, scalar and tensor primordial perturbations. With precise measurements of both sets of perturbations now apparently present, one could argue that it is time to turn the tables and solve the problem of super-Planckian quantum fields by adding complicated (i.e. multi-epoch) classical inflation trajectories.  

In this work we have shown that for $r \gtrsim 0.1$, there is no simple inflaton trajectory which allows for sub-Planckian, single field inflation, once phenomenological constraints on a single inflaton's trajectory are taken into account. The measurement of the spectral index over the 8 observed e-folds on the CMB alone implies
\begin{align}
\frac{\Delta \phi}{\mpl} \geq 1.00 \sqrt{\frac{r}{0.1}}, \label{eq:theboundagain}
\end{align}
for any theory with a spectral index describable by $\eta_V$ and $\epsilon_V$ over the measured 8 e-folds of inflation. As indicated in Eq.~(\ref{eq:theboundagain}), this constraint is  strong for any large value of $r \gtrsim 0.05$ and not tied to the exact value reported by the BICEP2 experiment. Further, this constraint will tighten with each future measured e-fold of small scale primordial perturbations, assuming those measurements do not show the power spectrum swiftly rising. Departures from this bound will require contriving (and arguably overfitting) single field models with spectral indices that depend on higher-order slow-roll parameters. 

If one wishes to construct a potential which remains barely sub-Planckian for $r \sim 0.1$, there are even more contrivances to be addressed.\footnote{And other bounds on small field inflation, see especially Ref.~\cite{Baumann:2011ws} and Ref.~\cite{tianjun}.} First, the inflaton trajectory must be non-monotonic in $\epsilon_V$. Second, the trajectory must shift from a curved trajectory typically associated with large field inflation to a very flat trajectory typically associated with inflection point models, in order to both fit cosmological perturbations over the observed 8 e-folds, and remain sub-Planckian over the remaining 50. Third, to remain sub-Planckian, these remaining 50 e-folds require that $\epsilon_V$ remain fixed and small. Even then, one must be careful not to run afoul of primordial black hole abundance constraints, which require the power spectrum remain less than $\sim 10^{-2}$ for $k \sim 0.2 - 10^{22} ~{\rm Mpc^{-1}}$. This last concern can be alleviated if a second epoch of inflation occurs at a smaller energy scale, because the results of Planck and BICEP2 do not forbid inflation at lower energy scales for the final 50 e-folds, during which the primordial modes are smaller than those we have measured at Hubble scale. Such models are typically called double or multiple inflation \cite{Silk:1986vc,Holman:1991ht,Adams:1997de,Burgess:2005sb}, often motivated by SUGRA inflation or hybrid potentials. Finally, one contrivance left unexplored in this work is that $\epsilon_V$ will have to increase again in order for inflation to end. 

Apparently for $r \gtrsim 0.1$ one cannot salvage a simple perturbative quantum field theoretic description of a sole inflaton which has order one couplings to other fields and a Planck scale cutoff. Still, multifield, sub-Planckian inflation leaves ample room for retreat if theorists wish to pursue models with large arrays of sub-Planckian ranging scalar fields \cite{Dimopoulos:2005ac,Ashoorioon:2009wa,Ashoorioon:2011ki,Ashoorioon:2014jja,Bachlechner:2014hsa}, or more recent proposals of a few scalar fields twisting inside a sub-Planckian patch of multi-scalar field space \cite{Harigaya:2014eta,McDonald:2014oza,Kim:2004rp,Kappl:2014lra}. On the other hand, proposals of shift-symmetric inflaton potentials transposed from UV completions of gravity \cite{Kaloper:2008fb,McAllister:2008hb,Flauger:2009ab,Kaloper:2011jz,Burgess:2011fa,Gur-Ari:2013sba,Long:2014dta,Baumann:2014nda,McAllister:2014mpa,Palti:2014kza,Hebecker:2014eua,Marchesano:2014mla,Kaloper:2014zba,Dine:2014hwa}, deserve more careful attention if the results of BICEP2 are confirmed. Rather than merely assuming the inflaton is weakly-coupled, we may wish to consider what Planck-scale symmetries may be manifest in a measurement of $r \gtrsim 0.1$.

\acknowledgements
We gratefully acknowledge useful discussions and correspondence with Prateek Agrawal, Nima Arkani-Hamed, Antonio Delgado, Austin Joyce, John Kearney, Tongyan Lin, Surjeet Rajendran, and Matt Reece. We thank Jessica Cook for insightful comments on early versions of this manuscript.

\appendix
\section{An Inflection Point Potential with $|\eta_V| > 1$}
\label{app:inflec}

In this appendix we demonstrate examples of Planck flat trajectories for which $|\eta_V-2\epsilon_V|$ becomes sufficiently large so as to avoid the bound of Antusch and Nolde. We also demonstrate how the use of these flat potentials to generate the remaining, observationally required e-foldings of inflation may lead to such a small $\epsilon_V$ that the power spectrum becomes unacceptably large.

Inflection point inflation is the flat inflation archetype. Because it is easy to realize, it has been studied extensively in string compactifications \cite{Kachru:2003sx,Kallosh:2004yh,BlancoPillado:2004ns} and particle physics models \cite{Lyth:1998xn,Allahverdi:2006iq,Mazumdar:2010sa,Allahverdi:2011su}. When two critical points of the potential degenerate, for example when $V_\phi \approx \mpl V_{\phi \phi}\approx0$, the leading term in the potential's Taylor expansion is cubic in $\phi$. The tiny plateau which results affords arbitrarily many e-foldings of inflation at the cost of tuned couplings in the potential \cite{Linde:2007jn}. Importantly, inflection points can serve as a dynamical attractor \cite{Itzhaki:2008hs,Downes:2012xb}; the nebulous fat tuning between high field values and small couplings inherent in initial conditions is traded for a potential with concrete tuning of couplings required to create the inflection point. For an in-depth discussion of these details see Refs.~\cite{Downes:2011gi,Downes:2012xb}.

Of course, both scenarios are possible. Planck fat inflation can precede a period of inflection point inflation. Such a transition to inflection point inflation can induce a rapid but large change in the slow-roll parameters. This ``hop'' away from the slow-roll attractor is induced by the smooth change from the power law trajectory of the fat inflation to the trajectory of the flat case. The magnitude of the hop --- and by extension an induced spike in the slow-roll parameters --- is directly related to the field velocity near the inflection point. In the present case, this velocity depends on both how the inflaton exits the fat stage of inflation. The tensor-to-scalar ratio induces a field velocity,
\begin{equation}\label{apx-eqn:rvel}\frac{1}{\mpl}\frac{d\phi}{dN} =  \sqrt{\frac{r}{8}}.\end{equation}
The ratio of couplings in the scalar potential determine how fast the kinetic energy of the inflaton redshifts with expansion. If either the velocity in \eqref{apx-eqn:rvel} is too large, or the redshifting is too slow, the field will overshoot the inflection point and a second phase of inflation generically will not occur. Fortunately for the present work, overshoot is not generic --- the inflection point possesses a basin of attraction in phase space. In other words, for a range of initial inflaton velocities, as the inflaton moves across the inflection point, inflation will occur. The closer to the boundary of this basin of attraction the inflaton is as it exits its fat phase, the larger the ``hop'' in the slow-roll parameters, see Refs.~\cite{Cicoli:2013oba,Jain:2008dw}. 

We demonstrate this in the specific case of a quartic potential tuned to have an inflection point
\begin{equation}\label{apx-eqn:V}V = V_0\left(\frac{1}{4}\phi^4 + \alpha\phi^3 + \lambda \phi \frac{27}{4}\alpha^4 + \mathcal{O}(\lambda)\right).\end{equation}
As with all inflection point models, this tuning can be made explicit by a small linear coupling which directly relates to the number of e-foldings generated.
Note also that $V_0$ is set by the potential during the fat phase. Let us now consider some particular examples. For $r=0.1$, \eqref{apx-eqn:rvel} gives an initial field velocity $\frac{1}{\mpl} d\phi/dN\approx-0.158$. The initial VEV of the inflaton will be represented by $\phi_0$.

For $\alpha=0.3~\mpl$ and $\phi_0=0.1~ \mpl$, one finds forty e-foldings of flat inflation when $\lambda\sim1.35\times 10^{-5}~\mpl^3$. In this case the power spectrum settles in to the fairly high value of $A_s\sim 4\times10^{-4}$ within four e-foldings. If fifty extra e-foldings of inflation were required, $\lambda\sim8\times10^{-6}~\mpl^3$ and $A_s \sim 8\times10^{-4}$. In these cases the average value of $|\eta_V-2\epsilon_V|$ is around $2$.

Decreasing $\alpha$ to $0.289 ~\mpl$ decreases the redshifting ability of the potential and so $\lambda$ must be smaller. To achieve fifty e-foldings, $\lambda\sim8\times10^{-6}~\mpl^3$ and $A_s\sim2\times 10^{-3}$. In this case the average value of $|\eta_V-2\epsilon_V|$ is around $1.4$.

Decreasing $\alpha$ further is difficult because we pass out of the basin of attraction of the inflection point. This demonstrates an important difficulty in realizing the fat-flat transition: the cubic coupling must be rather large. In any case, for $\alpha=0.275 ~\mpl$ and $\lambda = 2\times 10^{-6}~\mpl^3$, nearly fifty e-foldings are produced and $A_s\sim0.07$. For an extra sixty e-foldings, the power spectrum violates the black hole bound of Sec~\ref{sec:pbh}.

\begin{figure}[h!]
\includegraphics[scale=0.6]{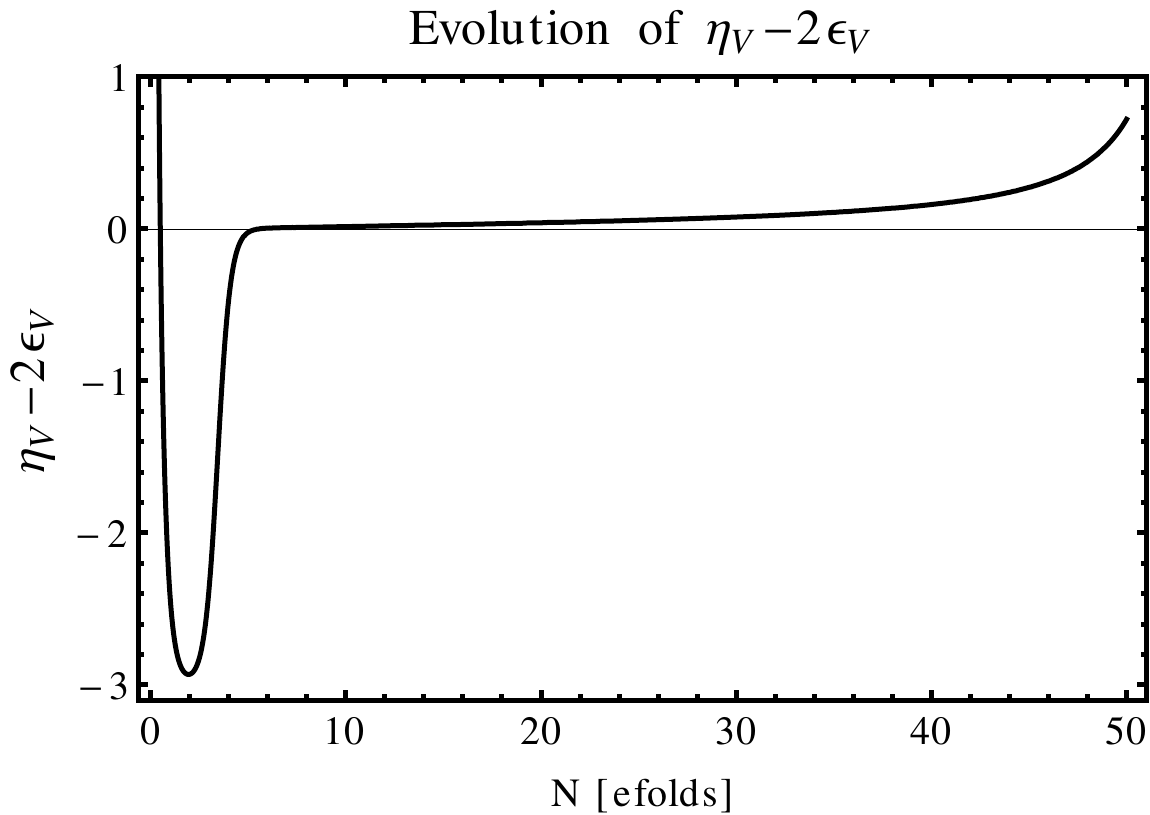}
\caption{The second logarithmic derivative of the Hubble parameter with respect to $N$, which corresponds to the slow-roll parameter combination $\eta_V - 2\epsilon_V$ used in the Antusch-Nolde discussion, is plotted as a function of e-folding. The inflaton evolves under the potential \eqref{apx-eqn:V}, with $\alpha=0.275~\mpl$ and $\lambda = 2\times10^{-6} ~\mpl^3$, generating fifty e-foldings of Planck flat inflation. Note the sharp, order unity drop at the onset of inflection point inflation.}
\label{apx-fig:AN}
\end{figure}

\begin{figure}[h!]
\includegraphics[scale=0.6]{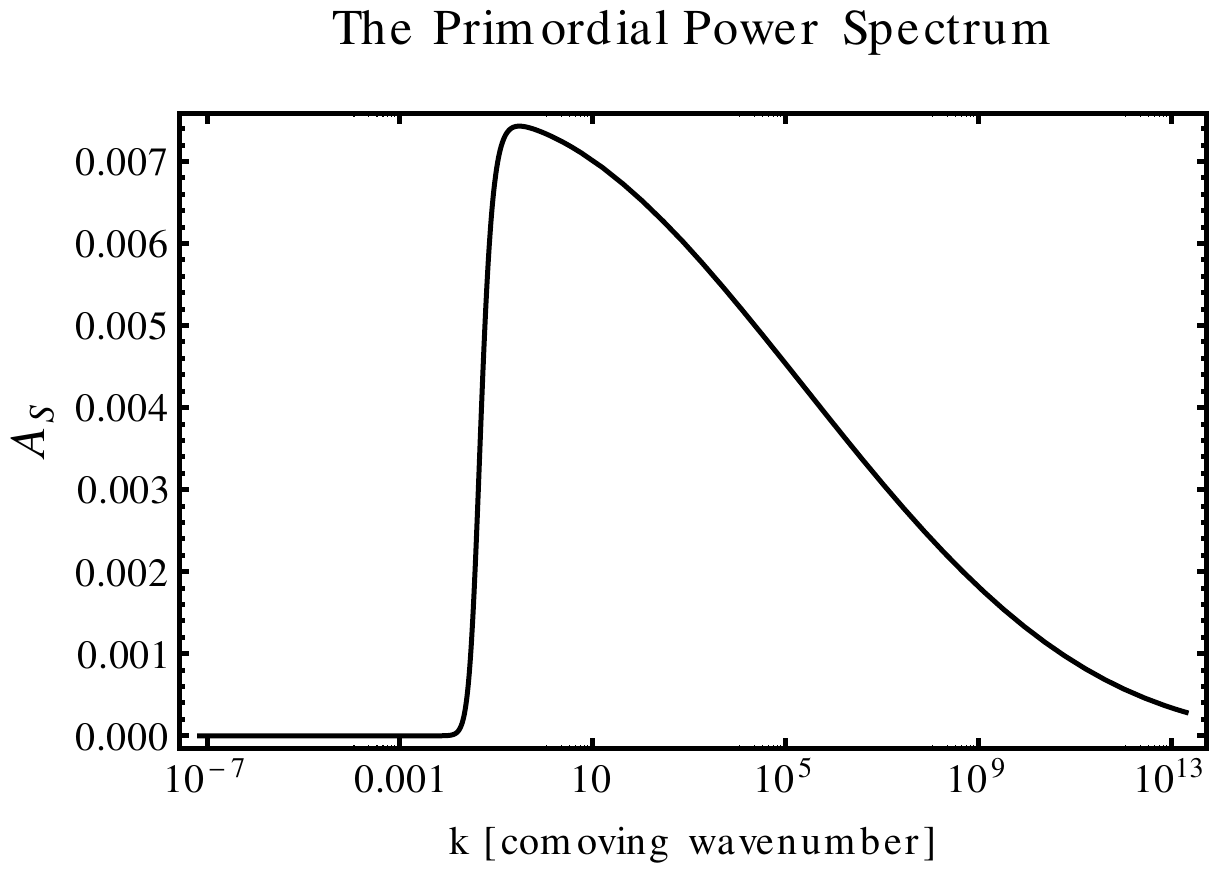}
\caption{The primordial, scalar power spectrum associated to the dynamics shown in Fig.~\ref{apx-fig:AN} is plotted. The inflaton evolves from the observed value of $10^{-9}$ under the potential \eqref{apx-eqn:V}, with $\alpha=0.275 ~\mpl$ and $\lambda = 2\times10^{-6} ~\mpl^3$, generating fifty e-foldings of Planck flat inflation. Note that it attains a value precariously close to the bound from primordial black hole generation, near $10^{-2}$.}
\label{apx-fig:PS}
\end{figure}

In Figs.~\ref{apx-fig:AN} and \ref{apx-fig:PS} we give a sample of the AN parameter $|\eta_V-2\epsilon_V|$ for $\alpha=0.275 ~\mpl$, $\lambda=2\times10^{-6} ~\mpl^3$ and its associated power spectrum.


In summary, the transition from fat to flat inflation is admissible provided one can engineer a potential to transition to inflection point inflation with about $\phi_0\sim 0.1 ~\mpl$ of field excursion remaining. This requirement is due to the field velocity required to generate the observed tensor modes during the Planck fat phase. As discussed in the main text, this is a highly nontrivial feat. Demanding a smaller $\phi_0$ pushes the system out of the basin of attraction for the inflection point. Being closer to the basin wall means a sharper decline in field velocity, which can bring the power spectrum fatally close to the black hole bound. Moreover, physical considerations which determine the number of required e-foldings, like the reheating temperature \cite{Allahverdi:2009rm}, will also be constrained by the black hole bound of the power spectrum. In summary, inflection point inflation fits, but just barely.

\section{More Potential Field Ranges for 8 e-folds}
\label{app:pots}
In this appendix we continue the program of Section \ref{sec:8e-fold}, examining simple sub-Planckian potentials which may achieve $r = 0.1,0.2$ and $n_s-1 = -0.04$ at $\phi_*$. We reiterate that any apparent deviation of the results from Eq.~\ref{eq:thebound} results from a spectral index increase between $\phi_*$ and $\phi_e$.
\subsection{Exponential potential}
Another simple potential, the exponential potential with $V(\phi) = \Lambda^4
\exp{\left(-\lambda \phi \right)}$, has $\epsilon_V = \lambda^2 /2$ and thus
$\lambda = \sqrt{r/8}$. Following the same procedure as in Section \ref{sec:8e-fold} gives
\begin{equation}
	N = 8 = \frac{1}{\mpl}\int_{\phi_e}^{\phi_*} \frac{\text{d}\phi}{\lambda} =
	\frac{1}{\lambda \mpl} (\phi_* - \phi_e) ,
	\label{}
\end{equation}
so $\Delta \phi = 8 \lambda \mpl = \mpl \sqrt{8r}$. For $r=0.1$, $\Delta \phi = 0.89 ~\mpl$, and
for $r=0.2$, $\Delta \phi = 1.26 ~ \mpl$.

\subsection{Starobinsky potential}
For Starobinsky inflation with a potential of the form $V(\phi) =
\frac{3}{4}\mu^2 \left(1 - \exp{\left(-\sqrt{\frac{2}{3}} \frac{\phi}{\mpl}\right)}
\right)^2$,
\begin{equation}
	\epsilon_V = \frac{4}{3 \left(\exp{\left(\sqrt{\frac{2}{3}}\frac{\phi}{\mpl}\right)} 
	- 1 \right)^2 } .
	\label{}
\end{equation}
Using $r$ to solve for $\phi_*$ as before gives
\begin{equation}
	\phi_* = \mpl \sqrt{\frac{3}{2}} \log{ \left(1+\frac{8}{\sqrt{3r}}\right)} ,
	\label{}
\end{equation}
and 
\begin{align}
	N = 8 &= \frac{1}{4 \mpl} \left(\sqrt{6} (\phi_*-\phi_e)+3
	e^{\sqrt{\frac{2}{3}}
	   \phi_e}-3 e^{\sqrt{\frac{2}{3}} \phi_*}\right)
	\label{}
\end{align}
Matching $r=0.1$ gives $\Delta \phi = 3.72-2.75 ~\mpl=0.97 ~\mpl$. 

\subsection{Symmetry breaking potential}
For a symmetry breaking potential of the form $V(\phi) = \Lambda^4
\left(1-\frac{\phi^2}{\mu^2}\right)^2$, 
\begin{equation}
	r  = \frac{128 \phi^2}{\left(\mu^2 - \phi^2\right)^2} .
	\label{}
\end{equation}
Since there are 2 parameters in $\epsilon_V$, we can calculate $n_s$:
\begin{equation}
	n_s = 1 + 2 \eta_V - 6 \epsilon_V = \frac{\mu ^4-2 \mu ^2 \left(\phi
	^2+4 \mpl^2\right)+\phi ^2 \left(\phi ^2-24 \mpl^2\right)}{\left(\mu
		   ^2-\phi ^2\right)^2}.
	\label{}
\end{equation}
For $r=0.1$, $\Delta \phi = 16.16 - 15.21 ~\mpl = 0.95 ~\mpl$.

\bibliographystyle{JHEP}

\bibliography{lythbh}

\end{document}